\documentclass[aps,prb,reprint,amsmath,amssymb,showpacs,showkeys %,groupedaddress,frontmatterverbose
]{revtex4-1}
\usepackage{graphicx}% Include figure files
\usepackage[colorlinks=true,allcolors=blue]{hyperref}%
\begin{document}
	
\title{Doped Silicon Quantum Dots as Sources of Coherent Surface Plasmons}
%\title{A Tunable Spherical Graphene Spaser}
	
\author{Sadreddin Behjati Ardakani}
\email[]{behjati@ee.sharif.ir}
\affiliation{Department of Electrical Engineering, Sharif University of Technology, Azadi Avenue, Tehran, Iran}
	
\author{Rahim Faez}
\email[]{faez@sharif.ir}
\affiliation{Department of Electrical Engineering, Sharif University of Technology, Azadi Avenue, Tehran, Iran}
	
\date{\today}
	
\begin{abstract}
In the present work, we propose using doped Silicon Quantum Dot (SiQD) as a source of coherent surface plasmons (SPASER). The possibility of spasing in single SiQD is investigated theoretically utilizing full quantum mechanical treatment. We show that spasing can take place in doped SiQDs whenever Quality factor of a plasmon mode exceeds some minimum value. The minimum value depends on size and doping concentration of SiQDs. It can be used to design an optimum structure as SPASER in silicon technologies. The condition on Quality factor is translated to a condition for radius and it is shown that for a given Localized Surface Plasmon (LSP) mode, the radius should be less than some critical value. This value only depends on mode index. The required relations for design purposes are derived and, as an example of feasibility of the approach, a SPASER is designed for mid infrared. Moreover, we propose a more applicable device by arranging an array of doped SiQDs on top of a graphene layer. Coupling the Surface Plasmon Polariton (SPP) modes of graphene with LSP modes of SiQDs causes outcoupling of an intense ultra narrow beam of coherent SPPs to be used in real probable applications.
\end{abstract}

\maketitle

\section{\label{sec:int}Introduction}
Coherent surface plasmon generation has been a topic of interest in plasmonics field since the introduction of SPASER (Surface Plasmon Amplification by Stimulated Emission of Radiation) in 2003 by Bergman and Stockman.\cite{bergman2003surface} The inventors showed that in some correctly tailored plasmonic structures, stimulated surface plasmons can be generated. SPASER can break the well known diffraction limit of light which is a bottleneck in laser technology. Since the introduction of SPASER, a number of researchers around the world have been focusing on its experimental and theoretical investigation. The first claimed experimental realization of SPASER dates back to 2009 in a paper reported by Noginov \textit{et al.}\cite{noginov2009demonstration} The authors experimentally showed that an aqueous solution of gold nanoparticles surrounded by dye-doped silica shells can behave like a SPASER. In 2010, Stockman treated the SPASER by a two level model and claimed that spasing action has only a quantum mechanical description.\cite{stockman2010spaser} In 2013, Zhong \textit{et al.} proposed a semiclassical approach for describing SPASER.\cite{zhong2013all} In the same year, Dorfman \textit{et al.} introduced a more controllable SPASER by its three level modeling.\cite{dorfman2013quantum} Many other works are found in the literature that concern different aspects of SPASER.\cite{andrianov2011forced,khurgin2012injection,li2013electric,parfenyev2014quantum,rupasinghe2014spaser,jayasekara2015multimode,totero2016energy,apalkov2014proposed,berman2013graphene,ardakani2017tunable,ardakani2017spaser}

In some ways, the SPASER acts like a laser. A device could be a SPASER if it consists of at least two main parts, a media that supports plasmonic modes and an active medium. Carrier down transition or LSP radiation in an initially pumped active medium is stimulated by inherently large positive feedback mechanism which is a consequence of LSP's intense near field. Outer energy supplies could be optical, electrical, chemical, or any other typical pumping systems that are used in laser, too.

Optical researches have interest to generate light from silicon. The indirect bandgap of silicon makes this material impossible to interact with photons directly. It has been shown that doped SiQDs have LSP modes.\cite{zhou2015comparative,kramer2015plasmonic,ni2017plasmonic} In the present paper, we claim that under some conditions a single doped SiQD could be a SPASER alone. Our assertion is based on the fact that a doped SiQD can take both the role of an active medium and an LSP supporter. The realized SiQD SPASER will have the advantage of its compatibility with silicon industries and capability of integration with silicon-based platforms.

\begin{figure}[tb]
	\centering
	\includegraphics[width=\linewidth]{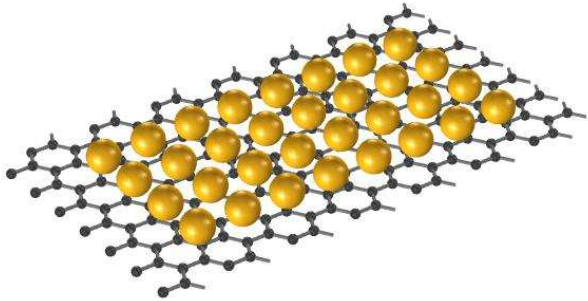}
	\caption{\label{fig:structure}(Color online) The structure used in this paper. SiQDs and graphene are sketched in gold and black, respectively. The graph is not shown in scale. All the structure is embodied in a SiO$_2$ matrix.}
\end{figure}

In this paper, we also use graphene as a media for propagating SPP modes. Graphene is a recently invented 2D material which is synthesized by a 2D arrangement of carbon atoms in a honeycomb lattice.\cite{novoselov2005two,geim2010rise} This material has excelent properties in multiple fields among which is plasmonics. Graphene has got a propagation length, lateral mode confinement, and lifetime, an order of magnitude larger than metals which are the most common materials in plasmonics area. In our work, graphene will be used to extract the energy of spasing LSP modes for external applications. Our proposed structure is shown in Fig.~\ref{fig:structure}
. The structure consists of an array of doped SiQDs on top of a graphene sheet which is embodied in a SiO$_2$ matrix. The graphene sheet is used for outcoupling of energy of coherent plasmons for real applications.

The paper is organized as follows. We first investigate the possibility of spasing in a single separated doped SiQD in the next three ongoing sections. So, in sections \ref{sec:Hlsp}, \ref{sec:active}, and \ref{sec:spasing}, with system we mean a single doped SiQD. Section \ref{sec:Hlsp} is about the subject of quantizing the LSP Hamiltonian of system. In section \ref{sec:active}, the required quantities in active medium are derived, and section \ref{sec:spasing} concludes the possibility of spasing in single doped SiQD. Finally, in section \ref{sec:res}, the whole structure is discussed and simulated numerically.  

\section{\label{sec:Hlsp}LSP Hamiltonian}
This section includes two parts. In the first part, LSP modes of a single doped SiQD is derived and the second part is devoted to calculating the quantized Hamiltonian of LSP field.

\subsection{\label{subsec:lsp}LSP modes of a single doped SiQD}
In this paper, Drude model is used for describing the dielectric constant of doped silicon. So it can be written as
\begin{equation}
	\epsilon_\mathrm{s}(\omega)=\epsilon_\infty-\frac{\omega_\mathrm{p}^2}{\omega(\omega+i\Gamma)},\label{eq:sipermi}
\end{equation}
where $\omega_\mathrm{p}$, $\epsilon_\infty$, and $\Gamma$ are Silicon's bulk plasma frequency and dielectric constant, and carrier's relaxation rate, respectively, and we have $\omega_\mathrm{p}^2=\frac{Ne^2}{\epsilon_0m^*}$, where $N$, $e$, $\epsilon_0$, and $m^*$ are doping concentration, elementary charge, vacuum permitivity, and carrier's effective mass, respectively. It is worth mentioning that, along the paper, the physicist's convention $\exp(-i\omega t)$ is used for all time dependencies.
 
LSP modes of a single doped SiQD with radius $r_0$ can be derived with a good accuracy by considering it as a quasi-electrostatic problem because discussed wavelengths are much larger than the maximun feature size of the system in hand. In the quasi-electrostatic limit, LSP modes are independent solutions of Laplace equation, $\nabla^2\Phi_{lm}(r,\theta,\phi)=0$, where $l$ and $m$ are mode indices. Utilizing the spherical symmetry of the problem, LSP modes could be written in the following form,
\begin{equation}
	\Phi_{lm}(r,\theta,\phi)=\left\lbrace
	\begin{array}{ll}
	\left(\frac{r}{r_0}\right)^l\,Y_l^m(\theta,\phi) & r\le r_0,\\
	\left(\frac{r}{r_0}\right)^{-(l+1)}\,Y_l^m(\theta,\phi) & r>r_0,
	\end{array}\right.
\end{equation}
where $Y_l^m$'s are spherical harmonics. Applying perpendicular electric displacement continuity condition at $r=r_0$ yields the following relation,
\begin{equation}
	S(\omega_{lm}-i\gamma'_{lm})=0,\label{eq:ceq}
\end{equation}
where $\omega_{lm}$ and $\gamma'_{lm}$ are $lm$'th mode's eigen-frequency and damping rate, respectively, and
\begin{equation}
S(x)=l\epsilon_\mathrm{s}(x)+(l+1)\epsilon_\mathrm{a}(x).
\end{equation}
$\epsilon_\mathrm{a}$ stands for ambient (where in our case is $\mathrm{SiO}_2$) dielectric constant. In the low damping regime, $\gamma'_{lm}\ll\omega_{lm}$, the solutions of complex equation Eq.~(\ref{eq:ceq}) are best approximated by the following two relations,
\begin{eqnarray}
	\omega_{lm}&\simeq&\sqrt{g_l(N)-\Gamma^2}\simeq\sqrt{g_l(N)},\label{eq:omega}\\
	\gamma'_{lm}&\simeq&\frac{g_l(N)\Gamma}{2(g_l(N)-\Gamma^2)}\simeq\frac{\Gamma}{2},\label{eq:gammap}
\end{eqnarray}
where
\begin{equation}
	g_l(N)=\frac{l\omega_\mathrm{p}^2}{l\epsilon_\infty+(l+1)\epsilon_\mathrm{a}}.
\end{equation}
Quality factor of modes is defined as $Q_{lm}=\omega_{lm}/2\gamma'_{lm}$ and a simple substitution leads to
\begin{equation}
	Q_{lm}\simeq\frac{\sqrt{g_l(N)}}{\Gamma(N)}\label{eq:Q}.
\end{equation}
In the above relation, we emphasize the doping dependency of carrier damping rate by explicitly writing it, but ,for simplicity, it is treated as a constant during this paper. Solutions of Eq.~(\ref{eq:ceq}) and consequently Eqs.(\ref{eq:omega}) and (\ref{eq:gammap}) apparently do not depend on the choice of mode index $m$. Therefore, for a fixed value of $l$, there exits $2m+1$ degenerate LSP modes. So for simplicity, throughout the paper, sometimes the index $m$ is omitted from the related quantities.

For the case of P-doped SiQD, using the parameter values listed in Table.~\ref{tab:param}
\begin{table}[bt]
	\renewcommand{\arraystretch}{1.3}
	\caption{\label{tab:param} Physical parameters which are used in this paper. In this table $m_e$ stands for electron mass.}
	\begin{tabular*}{\linewidth}{@{\extracolsep{\fill}}lcc}
		\hline\hline
		Quantity & Value & Unit\\
		\hline
		$\epsilon_\infty$ & 11.5 & --\\
		$\epsilon_\mathrm{a}$ & 4.5 & --\\
		$m^*_e$ & $0.26$ & $m_e$\\
		$m^*_h$ & $0.37$ & $m_e$\\
		$\Gamma$ & $4.5\times10^{11}-10^{13}$ & $\mathrm{rad/s}$\\
		$\Gamma_{pq}$ & $16$ & $\mathrm{meV}$\\
		$E_{\mathrm{G\,Si}}$ & 1.1 & $\mathrm{eV}$\\
		$E_{\mathrm{G\,SiO_2}}$ & 8.9 & $\mathrm{eV}$\\
		$\Delta E_\mathrm{c}$ & 3.4 & $\mathrm{eV}$\\
		$\Delta E_\mathrm{v}$ & 4.4 & $\mathrm{eV}$\\
		\hline\hline
	\end{tabular*}
\end{table},
the following approximate relations could be utilized for design purposes,
\begin{eqnarray}
	\omega_{lm}&=&1.11\times 10^5\;\sqrt{\frac{N[\mathrm{cm}^{-3}]}{16+4.5/l}},\label{eq:omegal}\\
	\lambda_{lm}[\mathrm{\mu m}]&=&1.7\times 10^{10}\;\sqrt{\frac{16+4.5/l}{N[\mathrm{cm}^{-3}]}}\label{eq:lambdal},
\end{eqnarray}
where $\lambda$ is LSP wavelength. Throughout the paper, all the quantities are assumed to be measured in SI unit system unless the unit is emphasized in a bracket next to the related quantity.

\subsection{\label{subsec:qlsp}Quantization of LSP Hamiltonian}
A general solution for electrostatic potential can be written as a linear combination of all LSP modes,
\begin{equation}
	\Phi(r,\theta,\phi)=\sum_{lm}C_{lm}\Phi_{lm}(r,\theta,\phi)\exp(-i\omega_{lm}t)+\mathrm{c.c.},\label{eq:phi}
\end{equation}
where $C_{lm}$'s are combination constants and c.c. stands for complex conjugate of previous terms. Electric field can also be written in the same way,
\begin{equation}
\mathbf{E}(r,\theta,\phi)=\sum_{lm}C_{lm}\mathbf{E}_{lm}(r,\theta,\phi)\exp(-i\omega_{lm}t)+\mathrm{c.c.},\label{eq:Emodes}
\end{equation}
where, $\mathbf{E}_{lm}=-\nabla\Phi_{lm}$.

For quantizing the LSP field, electrostatic energy has to be derived,\cite{landau2013electrodynamics}
\begin{equation}
	\mathcal{E}_\mathbf{e}=\frac{1}{2}\int\frac{\partial(\omega\epsilon)}{\partial\omega}\mathbf{E}\cdot\mathbf{E}\;\mathrm{d}^3r.\label{eq:energy}
\end{equation}
If Eq.~(\ref{eq:Emodes}) is substituted in Eq.~(\ref{eq:energy}), the following result is obtained,
\begin{eqnarray}
	\mathcal{E}_\mathrm{e}=\sum_{lml'm'}&C_{lm}&C_{l'm'}e^{ -i(\omega_{lm}+\omega_{l'm'})t}I_1\\
	+&C_{lm}&C^*_{l'm'}e^{-i(\omega_{lm}-\omega_{l'm'})t}I_2\\
	+&C^*_{lm}&C_{l'm'}e^{i(\omega_{lm}-\omega_{l'm'})t}I_3\\
	+&C^*_{lm}&C^*_{l'm'}e^{i(\omega_{lm}+\omega_{l'm'})t}I_4,
\end{eqnarray}
where
\begin{eqnarray}
	I_1&=&\frac{1}{2}\int\frac{\partial(\omega\epsilon)}{\partial\omega}\mathbf{E}_{lm}\cdot\mathbf{E}_{l'm'}\,\mathrm{d}^3r,\\
	I_2&=&\frac{1}{2}\int\frac{\partial(\omega\epsilon)}{\partial\omega}\mathbf{E}_{lm}\cdot\mathbf{E}^*_{l'm'}\,\mathrm{d}^3r,\\
	I_3&=&\frac{1}{2}\int\frac{\partial(\omega\epsilon)}{\partial\omega}\mathbf{E}^*_{lm}\cdot\mathbf{E}_{l'm'}\,\mathrm{d}^3r,\\
	I_4&=&\frac{1}{2}\int\frac{\partial(\omega\epsilon)}{\partial\omega}\mathbf{E}^*_{lm}\cdot\mathbf{E}^*_{l'm'}\,\mathrm{d}^3r.
	\end{eqnarray}
Tedious manipulations lead to the following results,
\begin{eqnarray}
	I_1&=&I_4=\frac{r_0}{2}(-1)^m\delta_{ll'}\delta_{m,-m'}A(\omega_{lm}),\label{eq:I1}\\
	I_2&=&I_3=\frac{r_0}{2}\delta_{ll'}\delta_{mm'}A(\omega_{lm})\label{eq:I2}.
\end{eqnarray}
where
\begin{equation}
	A(\omega_{lm})=\frac{\partial}{\partial\omega}\left\lbrace\omega S(\omega)\right\rbrace_{\omega=\omega_{lm}}.
\end{equation}
Substitution of Eqs.~(\ref{eq:I1}) and (\ref{eq:I2}) into Eq.~(\ref{eq:energy}) and averaging over time lead to
\begin{equation}
	\mathcal{E}_\mathbf{e}=\sum_{lm}\frac{r_0}{2}A(\omega_{lm})\left(C_{lm}^*C_{lm}+C_{lm}C_{lm}^*\right).
\end{equation}
Total energy is twice the average electrostatic energy so
\begin{equation}
	\mathcal{E}=\sum_{lm}r_0A(\omega_{lm})\left(C_{lm}^*C_{lm}+C_{lm}C_{lm}^*\right).
\end{equation}
LSP Field quantization is obtained by changing coefficients to the ladder operators by the following rule,
\begin{eqnarray}
	C_{lm}&\rightarrow&\gamma_{lm}\hat{a}_{lm},\label{eq:C}\\
	C^*_{lm}&\rightarrow&\gamma_{lm}\hat{a}^\dagger_{lm},\label{eq:Cdag}
\end{eqnarray}
where $\hat{a}_{lm}$ and $\hat{a}^\dagger_{lm}$ are annihilation and creation operators of $lm$'th mode which follow bosonic algebra and the following definition is made,
\begin{equation}
	\gamma_{lm}^2=\frac{\hbar\omega_{lm}}{2r_0A(\omega_{lm})}.
\end{equation}
The resulted quantized Hamiltonian is in the form of harmonic oscillator,
\begin{equation}
	\hat{H}_\mathrm{LSP}=\sum_{lm}\frac{\hbar\omega_{lm}}{2}(\hat{a}^\dagger_{lm}\hat{a}_{lm}+\hat{a}_{lm}\hat{a}^\dagger_{lm}).
\end{equation}
The operators of electrostatic potential and electric field also can be obtained by simply using Eqs.(\ref{eq:C}) and (\ref{eq:Cdag}) in Eqs.(\ref{eq:phi}) and (\ref{eq:Emodes}),
\begin{eqnarray}
	\hat{\Phi}(r,\theta,\phi;t)&=&\sum_{lm}\gamma_{lm}\Phi_{lm}\exp(-i\omega_{lm}t)\hat{a}_{lm}+\mathrm{H.c.},\\
	\hat{\mathbf{E}}(r,\theta,\phi;t)&=&\sum_{lm}\gamma_{lm}\mathbf{E}_{lm}\exp(-i\omega_{lm}t)\hat{a}_{lm}+\mathrm{H.c.},\label{eq:Eop}
\end{eqnarray}
where H.c. means hermitian conjugate of previous terms. Until now, the results are general and do not depend on what model we use for describing permittivity of doped Si, but if the specific form of $\epsilon_\mathrm{s}(\omega)$, from Eq.~(\ref{eq:sipermi}), is substituted in $\gamma_{lm}$ the following result is obtained,
\begin{equation}
	\gamma_{lm}^2=\frac{\hbar}{4\epsilon_0\omega_p^2}\frac{\omega_{lm}^3}{r_0l}.
\end{equation}

\section{\label{sec:active}S\lowercase{i}QD as an Active Medium}
According to Table.~\ref{tab:param}, the barrier height in SiQD is large enough to approximate it as an infinite well spherical quantum dot. Solving Shr\"{o}dinger equation leads to the following wavefunctions,\cite{ardakani2017tunable}
\begin{equation}
\psi_{kns}(r,\theta,\phi)=\left\lbrace\begin{array}{ll}
B_{nk}\;j_n\left(\frac{x_{nk}}{r_0}r\right)Y_n^s(\theta,\phi) &r\le r_0,\\
0 &r>r_0,
\end{array}\right.\label{eq:wfc}
\end{equation}
where $x_{nk}$ is the $k$'th zero of the $n$'th order spherical Bessel function of the first kind, $j_n$, and
\begin{equation}
B_{nk}=\left(\frac{2}{r_0^3[j_{n+1}(x_{nk})]^2}\right)^{0.5}
\end{equation}
is a normalization constant. Energy eigenvalues can be calculated by the following formula,
\begin{equation}
E_{kns}=\frac{\hbar^2x_{nk}^2}{2m^*r_0^2}.\label{eq:E}
\end{equation}
The important quantities in active medium, which are involved in interaction Hamiltonian, are dipole matrix elements. It is straight forward to compute these quantities by using Eq.~(\ref{eq:wfc}). Calculation leads to the following relation for dipole matrix element $\mathbf{d}_{pq}$ between general levels $p=nks$ and $q=n'k's'$,
\begin{equation}
\mathbf{d}_{pq}=-\hat{\mathbf{r}}\delta_{nn'}\delta_{ss'}er_0f_{nkk'},\label{eq:acdp}
\end{equation}
where $f$ is a dimensionless parameter which is independent of the choice of geometry,\cite{ardakani2017tunable}
\begin{equation}
f_{nkk'}=\frac{2}{j_{n+1}(x_{nk})j_{n+1}(x_{nk'})}\int_0^1r^3j_n(x_{nk}r)j_{n}(x_{nk'}r)\,\mathrm{d}r.
\end{equation}
For design purposes, $r_0$ should be designed such that there is a $p\leftrightarrow q$ transition that is getting close to resonance with one of LSP modes, say $l=L$. The numerical calculations in the rest of the paper are based on $\left|\psi_{10s}\right>\leftrightarrow\left|\psi_{20s'}\right>$ transitions. The result of calculation of $f_{012}$ for this case becomes 0.180128. In this case if zero of energy is chosen such that to coincide with the middle of two levels in resonance, the approximate active medium Hamiltonian in the second quantized form would be\cite{walls2007quantum,scully1997quantum}
\begin{equation}
H_\mathrm{a}=\frac{\hbar\omega_0}{2}\hat{\sigma}_z,
\end{equation}
where
\begin{eqnarray}
\hat{\sigma}_z &=& \hat{c}_q^\dagger\hat{c}_q-\hat{c}_p^\dagger\hat{c}_p,\\
\omega_0 &=& \frac{E_q-E_p}{\hbar}.
\end{eqnarray}
In the above relations $\hat{\sigma}_z$, $\hat{c}_j$, $\hat{c}_j^\dagger$ are  pseudo-spin operator, fermionic annihilation and creation operators of $j$'th level (for $j=p,q$), respectively. Furthermore, it is assumed that $E_q>E_p$, without loss of generality.

\section{\label{sec:spasing}Spasing in a single doped S\lowercase{i}QD}
For the rest of the paper, we suppose that only those LSP modes with $l=L$ are in near resonance with $\hbar\omega_0$ transition with $\mathbf{d}_{pq}=d\,\hat{\mathbf{r}}$ and so the interaction Hamiltonian can be written as,\cite{walls2007quantum,scully1997quantum}
\begin{equation}
H_\mathrm{I}=\sum_{m=-L}^L\hbar\Omega_{Lm}\left(\hat{a}_{Lm}^\dagger\hat{\sigma}_-+\hat{\sigma}_+\hat{a}_{Lm}\right),
\end{equation}
where $\Omega_{Lm}=-\mathbf{d}_{pq}\cdot\mathbf{E}_{Lm}/\hbar$ is Rabi frequency and
\begin{eqnarray}
	\hat{\sigma}_-&=&\hat{c}_p^\dagger\hat{c}_q,\\
	\hat{\sigma}_+&=&\hat{c}_q^\dagger\hat{c}_p,
\end{eqnarray}
are ladder operators. Rabi frequency is a measure of interaction strength.
%\begin{equation}
%	\Omega_{lm}=-\frac{\mathbf{d}\cdot\mathbf{E}_{lm}}{\hbar}.
%\end{equation}
Using electric field from Eq.~(\ref{eq:Eop}) in Rabi frequency results the following relation, 
\begin{equation}
	\Omega_{Lm}=\frac{d\gamma_L}{r_0\hbar}Y_L^m(\theta,\phi)\label{eq:rabi}.
\end{equation}
The condition for spasing has been derived in Ref.~\onlinecite{stockman2010spaser} and for near resonance circumstances it simplifies to the following inequality condition,
\begin{equation}
	\sum_{m=-L}^L\left|\Omega_{Lm}\right|^2\ge\gamma'_L\Gamma_{pq},\label{eq:spscond}
\end{equation}
where $\Gamma_{pq}$ is polarization relaxation rate of $\hbar\omega_0$ transition. By substituting Eq.~(\ref{eq:rabi}) into Eq.~(\ref{eq:spscond}) and some computations the following relation is obtained,\cite{ardakani2017tunable}
%\begin{equation}
%	\frac{d^2\omega_1^3}{4V_\mathrm{QD}\hbar\epsilon_0\omega_p^3}\ge\gamma'_1\Gamma_{pq}.
%\end{equation}
\begin{equation}
\frac{\pi d^2\omega_L^3}{3V_\mathrm{QD}\hbar\epsilon_0\omega_p^2}\ge\gamma'_L\Gamma_{pq}.
\end{equation}
where $V_\mathrm{QD}$ is SiQD's volume. This relation can be translated to the neat form, $Q_L\ge Q_{L\mathrm{min}}$, where
%\begin{equation}
%	Q_{1\mathrm{min}}=\frac{2\hbar V_\mathrm{QD}\epsilon_0\omega_\mathrm{p}^2\Gamma_{pq}}{d^2\omega_1^2}.
%\end{equation}
\begin{equation}
Q_{L\mathrm{min}}=\frac{3\hbar V_\mathrm{QD}\epsilon_0\omega_\mathrm{p}^2\Gamma_{pq}L}{2\pi d^2\omega_L^2(2L+1)}.\label{eq:Qmin}
\end{equation}
It is seen that $Q_{L\mathrm{min}}$ is capable of engineering because it depends on SiQD size and doping concentration through $\omega_\mathrm{p}$. This result can be compared with Ref.~\onlinecite{ardakani2017tunable} where the SPASER is made by a  spherical graphene shell and an array of Quantum Dots (QDs) around it. The present SPASER has two main advantage over it, which are integration capability with silicon platforms and simplicity of fabrication.\cite{ni2017plasmonic} Design parameters of the present work include SiQD's size and doping concentration in comparison to that work which were QD's size and array density and graphene sphere's radius. For Numerical purposes, it can be shown that
\begin{equation}
	Q_{L\mathrm{min}}=5.451\times 10^7\frac{16L+4.5}{2L+1}\,r_0.
\end{equation}
From the aforementioned relations, a simpler design condition is obtained, $r_0\le r_{0\mathrm{c}}$, where
\begin{equation}
	r_{0\mathrm{c}}=2.283\times 10^{-8}\left(\frac{2L+1}{16L+4.5}\right)^{1/3}.
\end{equation}
%\begin{equation}
%\frac{r_0}{N}<\frac{3m^*f^2e^4}{8h\Gamma_{pq}\Gamma(\epsilon_\infty+2\epsilon_\mathrm{a})^2},	
%\end{equation}
%\begin{equation}
%\frac{r_0^2}{N}<\frac{f^4e^6L(2L+1)^2}{4\hbar^2\epsilon_0^3m^*\Gamma_{pq}^2\Gamma^2[L\epsilon_\infty+(L+1)\epsilon_\mathrm{a}]^3},	
%\end{equation}
%\begin{equation}
%	r_{0\mathrm{c}}(\omega_L)=\left(\frac{f^2e^4(2L+1)}{2\hbar\epsilon_0\Gamma_{pq}\Gamma[L\epsilon_\infty+(L+1)\epsilon_\mathrm{a}]}\right)\times\omega_L,\label{eq:r0c}
%\end{equation}
%and $f$ is the $f$-number for $\hbar\omega_0$ transition. So the required relations for design purpose have already been obtained, Eq.~(\ref{eq:omegal}), Eq.~(\ref{eq:lambdal}), Eq.~(\ref{eq:E}), and Eq.~(\ref{eq:r0c}).

So, the required relations for design have already been derived and for compactness are listed below,
\begin{eqnarray}
	N[\mathrm{cm}^{-3}]&=&2.89\times 10^{20}\,\frac{16+4.5/L}{(\lambda_L[\mu\mathrm{m}])^2},\\
	r_0[\mathrm{nm}]&=&1.871\sqrt{\lambda_L[\mu\mathrm{m}]},\\
	r_{0\mathrm{c}}[\mathrm{nm}]&=&22.83\left(\frac{2L+1}{16L+4.5}\right)^{1/3}.
\end{eqnarray}
For an example we could have a SPASER that radiates at $\lambda=3\,\mathrm{\mu m}$ by $L=1$, $N=6.58\times10^{20}\,\mathrm{cm}^{-3}$, $r_0=3.24\,\mathrm{nm}$, and $r_{0\mathrm{c}}=12.03\,\mathrm{nm}$, which are typical values in fabrication technologies.

\section{\label{sec:res}Numerical Results and Discussions}
As it is mentioned in section \ref{sec:int}, the whole structure consists of an array of doped SiQDs on top of a sheet of graphene. Until now, the response of a single separated doped SiQD is assessed. Arranging doped SiQDs in an array and also coupling to the SPP modes of graphene sheet cause the LSP eigen-frequencies to be split and shifted. Furthermore, we expect a large field enhancement in the gap between SiQDs and graphene due to the constructive interference. SPP modes have a traveling wave character. So, this specific structure helps outcoupling of the spasing mode energy for real applications, just the same as what is done in laser by using imperfect mirrors for cavities.

\begin{figure}[tb]
	\centering
	\includegraphics[width=\linewidth]{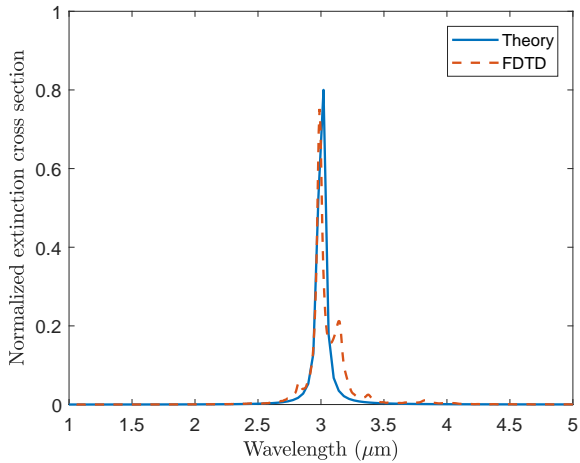}
	\caption{\label{fig:crossSection}(Color online) Extinction cross section of a single doped SiQD normalized to its geometrical cross section, $\pi r_0^2$.}
\end{figure}

Fig.~\ref{fig:crossSection}
compares the extinction cross section of a single separated doped SiQD with $r_0=3.24\,\mathrm{nm}$ and $N=6.58\times10^{20}\,\mathrm{cm}^{-3}$ calculated from theory and Finite Difference Time Domain (FDTD) simulations. Theoretical extinction has been derived by the following formula,\cite{maier2007plasmonics,bohren2008absorption}
\begin{equation}
	C_\mathrm{ext}=C_\mathrm{sca}+C_\mathrm{abs},
\end{equation}
where scattering and absorption cross sections are calculated by
\begin{eqnarray}
	C_\mathrm{sca} &=& \frac{8\pi}{3}k_L^4r_0^6\left|\frac{\epsilon_\mathrm{s}-\epsilon_\mathrm{a}}{\epsilon_\mathrm{s}+2\epsilon_\mathrm{a}}\right|^2,\\
	C_\mathrm{abs} &=&4\pi k_Lr_0^3\mathrm{Im}\left[\frac{\epsilon_\mathrm{s}-\epsilon_\mathrm{a}}{\epsilon_\mathrm{s}+2\epsilon_\mathrm{a}}\right],
\end{eqnarray}
and where $k_L$ is ambient wavenumber of $L$'th mode. The good agreement between theory and simulation verifies the accuracy of our analysis.

\begin{figure}[tb]
	\centering
	\includegraphics[width=\linewidth]{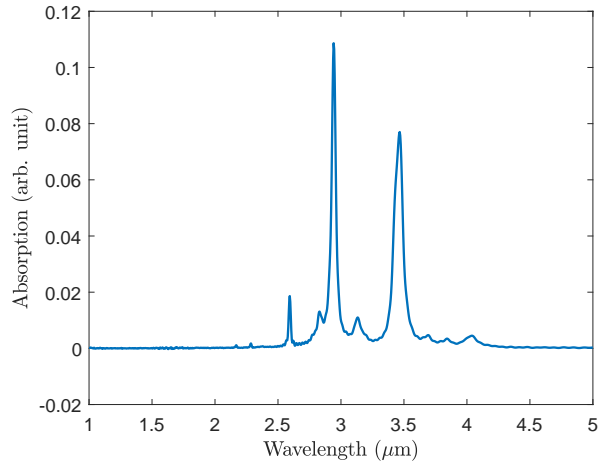}
	\caption{\label{fig:abs}(Color online) Absorption spectrum of the SiQD array which sits in a distance of $4\,\mathrm{nm}$ on top of a graphene sheet with $E_F=0.8\,\mathrm{eV}$ and scattering rate of $0.11\,\mathrm{meV}$.}
\end{figure}

\begin{figure*}[tb]
	\centering
	\includegraphics[width=.45\linewidth]{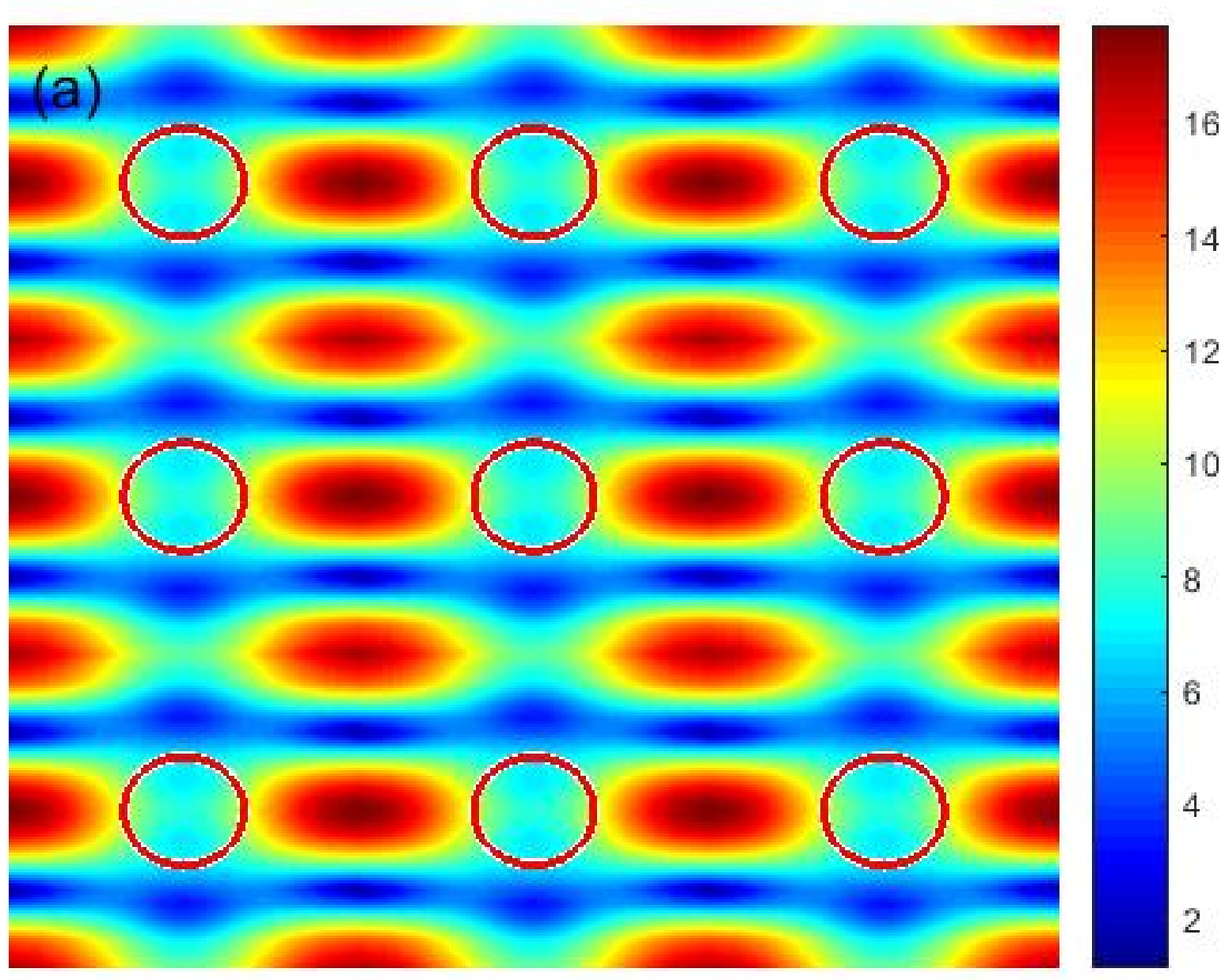}
	\includegraphics[width=.45\linewidth]{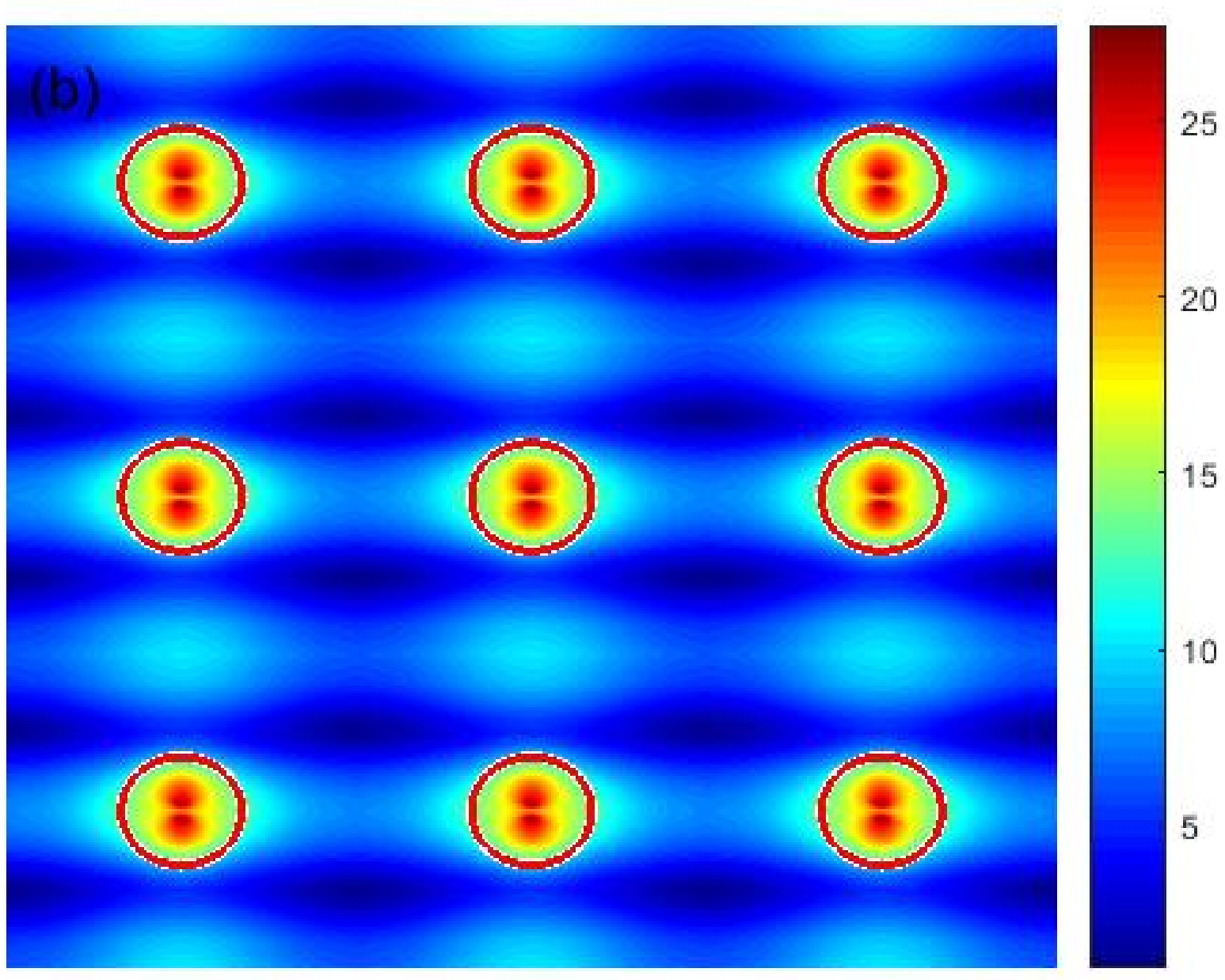}	\caption{\label{fig:profs}(Color online) Electric field magnitude distribution on graphene for (a) the first and (b) the second resonance occurred in Fig.~\ref{fig:abs}. Apparently in the case of (a) the energy of LSP mode has outcoupled from SiQDs and starts to propagate along graphene. For visual convenience, outlines of SiQDs have been sketched in red.}
\end{figure*}

The whole structure is illuminated by a normally incident plane wave. The SPP modes of graphene can not couple to this wave directly. But, for the efficient coupling between SPPs and the spasing LSP, array's period ($\Lambda$) could be chosen such that the first order diffracted wave coincides with an SPP mode in LSP wavelength, $2\pi/\Lambda=k_\mathrm{SPP}(\lambda_L)$, where $k_\mathrm{SPP}(\lambda_L)$ satisfies the following implicit equation,
\begin{equation}
	\frac{1}{\sqrt{k_\mathrm{SPP}^2-\epsilon_\mathrm{a}k_L^2}}=\frac{\sigma_\mathrm{2D}(\omega_L)}{2i\omega_L\epsilon_\mathrm{a}\epsilon_0}.
\end{equation}
The calculated period for $\lambda_L=3\,\mathrm{\mu m}$ is $\Lambda=18.833\,\mathrm{nm}$. The simulated absorption spectra for this choose of parameters is shown in Fig.~\ref{fig:abs}
.
This figure exhibits two main peaks. The coupling of SPP and LSP is responsible for the sharper peak around $2.94\,\mathrm{\mu m}$. The second peak around $3.47\,\mathrm{\mu m}$ indicates a higher order LSP mode. It can be seen from this figure that the $3\,\mathrm{\mu m}$ designed wavelength for single SiQD is slightly blue-shifted due to the coupling nature. The electric field distribution for SPP and LSP peaks are shown in Fig.~\ref{fig:profs}(a) and (b),
respectively. Apparently, LSP mode has been coupled to a SPP mode in Fig.~\ref{fig:profs}(a) but the distribution in Fig.~\ref{fig:profs}(b) is mostly localized.
 
\section{Conclusion}
In summary, we have claimed that correctly designed doped SiQDs can have some spasing modes. The inspiration behind that is the existence of LSP modes in doped SiQDs. Because a single doped SiQD can take both the role of active medium and LSP supporter, there must be some conditions under which the system can spase. We have analyzed the structure thoroughly using full quantum mechanical treatment and derived the required conditions for spasing. We have shown that by appropriately choosing the SiQD's size and doping concentration a SPASER could be designed for a given wavelength. We have also explained how to choose the array period for efficient outcoupling of spasing energy from SiQDs. During the paper, a SPASER  has been designed at $\lambda=3\,\mathrm{\mu m}$ and the results have been verified by using the FDTD simulations.

\bibliography{references}

\end{document}